\documentclass[11pt]{llncs}
\usepackage[letterpaper,hmargin=1in,vmargin=1.25in]{geometry}

\usepackage{latexsym}
\usepackage{amssymb}
\usepackage{times}
\usepackage{fullpage}
\usepackage{txfonts}
\usepackage{times,mathptm}

\newcommand{\ignore}[1]{}

\newcommand{\remove}[1]{}

\newlength{\saveparindent}
\newlength{\saveparskip}
\begin{document}

\title{(Unconditional) Secure Multiparty Computation with Man-in-the-middle Attacks\\
}

\author{{\sc Shailesh Vaya}\thanks{This work was done while the author was at the Department of Computer Science and Engineering, Indian Institute of Technology Madras, Chennai.}}

\institute{
\email{E-mail: vaya@cse.iitm.ac.in}\\
Department of Computer Science and Engineering\\
Indian Institute of Technology, Madras\\
Chennai, India - 600036.
}

\date{\mbox{ }}

\maketitle

\vspace{-0.3in}

\begin{abstract}
  In secure multi-party computation $n$ parties jointly evaluate an $n$-variate function $f$ in the presence of an adversary which can corrupt up till $t$ parties. All honest parties are required to receive their correct output values, irrespective of how the corrupted parties under the control of the adversary behave. The adversary should not be able to learn anything more about the input values of the honest parties, then what can be inferred from the input and output values of the corrupted parties and structure of the function.

  Almost all the works that have appeared in the literature so far assume the presence of authenticated channels between the parties. This assumption is far from realistic. Two directions of research have been borne from relaxing this (strong) assumption: (a) The adversary is virtually omnipotent and can control all the communication channels in the network, (b) Only a partially connected topology of authenticated channels is guaranteed and adversary controls a subset of the communication channels in the network.

  This work introduces a new setting for (unconditional) secure multiparty computation problem which is an interesting intermediate model with respect to the above well studied models from the literature (by sharing a salient feature from both the above models). We consider the problem of (unconditional) secure multi-party computation when 'some' of the communication channels connecting the parties can be corrupted passively as well as actively. We model communication channels as entities just like parties and consider a few different types of channels, namely fully secure channels, authenticated but eavesdroppable channels, partially tamperable channels and fully tamperable channels. For this setting, some honest parties may be connected to several other honest parties via corrupted channels and may not be able to authentically/privately communicate with them. Such parties may not be assured the canonical guarantees of correctness or privacy. Honest parties which are not guaranteed correctness or privacy properties are called sacrificed, as is done for the notion of almost everywhere secure computation (above model (b)). We present appropriate definitions of security for this new intermediate model of secure computation for the stand alone setting. We show how to adapt protocols for (unconditional) secure multiparty computation to realize the definitions and also argue the tightness of the results achieved by us.
\end{abstract}

\noindent {Keyword(s):} Secure multiparty computation, Byzantine corruption, Almost everywhere secure multiparty computation, Man-in-the-middle attack, Corruption of channels, Simulation paradigm, Simulation, Input indistinguishability.

\section{Introduction}
\label{sec:intro}
  In secure multiparty computation a set of $n$ parties jointly compute some function $f$ of their inputs in a secure way, in the presence of an adversary controlling a subset of $t$ parties. There are two canonical guarantees of the interactive computation process: (1) Correctness (2) Privacy. The correctness guarantee is that the correct output value is received by all the honest parties. The privacy guarantee is that irrespective of how the corrupted parties behave the adversary learns nothing more about the private inputs of the honest parties, then what can be inferred from the (initial/committed) input values of the corrupted parties, the output value and the structure of the function $f$.

  Since the seminal works on secure multiparty computation, \cite{Yao82}, \cite{GMW87}, \cite{BGW88} and \cite{CCD88} this area has been heavily explored. The one common assumption which almost all these works make is the availability of authentic/secure channels between every pair of parties. The reason for this assumption has been the belief that the communication channels can be easily realized through physical infrastructures like LAN, WAN, fibre optic cables etc. which provide the basic security features or by cryptographic schemes like PKC. Secondly, it seems to have been assumed that no meaningful guarantees can be achieved if basic reliable communication is not assured to the parties.

  The former belief is not well founded, because physical channels, though realistic, usually cannot assure reliability, let alone secrecy, in the presence of an active intruder. Secondly, it is not realistic to assume the availability of such channels between every pair of parties in the network because it imposes a very high requirement on the communication infrastructure. This becomes specially relevant in the information theoretic setting when PKI, digital signatures become irrelevant. Thus, three natural settings arise from questioning these assumptions: What meaningful security guarantees can be achieved (1) when no authentication mechanism is available (2) when only a partially connected topology of secure channels is available (3) when the adversary can corrupt and control the communication channels connecting the parties to various extents.

  \cite{BCLPR} consider this very weak model, of type (1), when the adversary can fully control the communication between the parties. For such a feeble model, not much can be guaranteed because the adversary can complete disrupt any computation process. Yet theoretically this is an interesting model to consider and the results given in \cite{BCLPR} have an interesting flavor. It is shown that the adversaries strategy can be restricted to either message relaying or conducting independent executions with disjoint subsets of honest parties, and the only dependence that the adversary can achieve, between these executions, is by running these executions sequentially and choosing its input in second execution after receiving its input from the first.

  \cite{V07}, \cite{GO08} consider the second model when only a partially connected network of secure channels is available to the parties, called almost everywhere secure computation. In contrast to \cite{BCLPR}, this model is theoretically and practically interesting for the case of information theoretic adversary only. On the other hand, although the adversary considered in \cite{BCLPR} is given too much power to assure any meaningful security guarantees for practical purposes, a.e.s.c. actually assures meaningful security guarantees to a large number of honest parties, if not all the honest parties (communistic settings are one example where they become relevant). Recall that the canonical guarantees of correctness and privacy has to be sometimes "sacrificed" for some honest parties that are surrounded by bad neighborhood of maliciously corrupted parties. Thus, depending on the statistical guarantees about the subset of maliciously corrupted parties, a.e.s.c. makes only statistical guarantees about the subsets of honest parties for which correctness and/or privacy properties are assured, for special families of incomplete networks.

  We consider the intermediate setting (c) when an "information theoretic" adversary (which renders PKI based solutions useless) can partially control a subset of communication channels in various ways. Depending on the actual subset of communication channels corrupted by the adversary and the type of corruptions of the channels, not all honest parties may be able to achieve the guarantees of correctness and privacy just as in \cite{V07}, \cite{GO08}. For example, if a fairly large number of communication channels connecting an honest party are totally corrupted, then this party may not be able to commit to its intended input value or guaranteed to receive its correct output value. On the other hand, if a large number of communication channels connecting to this honest party are authentic but eavesdroppable, then the privacy of its input value or that of its output cannot be guaranteed. Thus the canonical guarantees of correctness or privacy may have to be dropped for some of the honest parties. We present results for the new model of secure computation, in the framework of almost everywhere secure computation.

  Results in \cite{GO08} for a.e.s.c. hold for the case of honest-but-curious type of passive corruptions (The reader is referred to \cite{V10} for related exposition). However, theoretically meaningful and practically relevant realization of this model is for the case of malicious corruptions. Handling malicious corruptions requires a novel approach for formulating the definitions of security and in particular, the privacy property of protocols because we find that for this model there is no satisfactory way to apply the Trusted third party paradigm. In \cite{V07}, an appropriate definitional framework was proposed to realize almost everywhere secure computation on special types of incomplete networks, for handling the case of malicious adversary.

  In this work, we refine the definitional approach proposed in \cite{V07} further, to present results for the new model: (Unconditional) secure multi-party computation with man-in-the-middle attacks. We define the Correctness and Privacy properties for the new model separately. Thus, our definition of security for the new model has the flexibility to guarantee correctness property for a certain subset of honest parties and privacy property to a different subset of honest parties. Exactly which subsets of honest parties are guaranteed the Correctness or Privacy properties, of course depends on the subset of corrupted parties and channels and how the parties and communication channels between the parties are corrupted by the adversary i.e., the specific element of the adversary structure. We capture the various types of man-in-the-middle attacks on the communication channels, by associating with each channel an ideal description of its behavior (and in particular interaction with the adversary) under different types of attacks. Finally, we show how to realize our definitions of security for the new model by adapting standard protocols for unconditional secure multi-party computation.

  Lastly, from a practical point of view many different types of man-in-middle attacks may be considered in realistic scenarios. Computer networks often face the threat of the following type of security attack. An intruder inserts itself between two communicating parties, both of which believe that they are talking to each other, while the attacker deletes, modifies or simply eavesdrops the messages exchanged between the parties. More generally, such an attacker can carry out the attack in a coordinated fashion and sabotage a large number of communication channels of the network. In particular, the attacker may use the messages received on one communication channel to modify or inject new messages on a different channel. At the same time interesting variations exist for the lower level protocols, for protecting integrity and privacy of the messages communicated over the channels. They give rise to interesting combinations of channels and man-in-middle attacks in this work.

\subsection{Related works}
\label{relatedwork}
  The notion of almost everywhere secure computation for incomplete network and an overall approach to realize it, was presented in \cite{GO08}, by Garay and Ostrovsky and Vaya in \cite{V07}. In \cite{GO08}, the authors present Input indistinguishability type definition of privacy for almost everywhere secure computation. The input indistinguishability type definitional approach was first proposed in \cite{KKMO94} in a different context. A hybrid argument was given to realize this definition for honest-but-curious type {\it passive corruptions} in \cite{GO08},\cite{KKMO94}.

  Assuming that strictly more than $\lfloor \frac{2*n}{3} \rfloor$ parties are honest, it has been shown that it is possible to securely compute any $n$-variate function, \cite{BGW88}, \cite{CCD88} for the information theoretic regime. In the computational model, the results have been given in \cite{GMW87}, \cite{Yao82}.

  The trusted third party paradigm was proposed in \cite{GMW87} (It has been extended to propose universal composability framework in its most general form).

\subsection{Organization of the paper}
\label{organize}
  In Section \ref{Model}, we present preliminaries requisite for the presentation of our results. In Section \ref{sec:secmpc}, we present a complete definition of security for the stand alone setting. In Section \ref{interesting}, we discuss the application of this definition of security to a few interesting settings of unconditional secure multiparty computation.

\section{Preliminaries}
\label{Model}
  We first review some standard notations and terminologies for (unconditional) secure multiparty computation. This is followed by formal definitions of security for the vanilla setting.
\begin{definition}
\label{negligible}
  A function $\delta: N \rightarrow [0,1]$ is called {\it negligible} if for all $c > 0$ and for all large enough $k \in N$, we have $\delta(k) < k^{-c}$.
\end{definition}
\begin{definition}
\label{ensemble}
  A {\it distribution ensemble} $X = \{X(k,a)\}_{k \in N,a \in \{0,1\}^*}$ is an infinite set of probability distributions, where $X(k,a)$ is associated with each $k \in N$ and $a \in \{0,1\}^{*}$. A distribution ensemble is called binary if it consists only of distributions over $\{0,1\}$.
\end{definition}
\begin{definition}
\label{App:statisticalindistinguishability}
  Distribution ensembles $X$ and $Y$ are called {\it statistically indistinguishable} if for all sufficiently large $k$ and $a$, $SD(X,Y) = \frac{1}{2} \sum_a |Prob(X=a) - Prob(Y=b)| < \delta(k)$ for $\delta(.)$ a negligible function.
\end{definition}

  $Power-Set(F)$ refers to the set of all subsets of a set $F$. A mixed network $N$ is referred to as a triplet $(V,E,E_d)$, where $V$ refers to a set of vertices, and $E$ refers to the set of undirected edges and $E_d$ refers to the set of directed edges.

\subsection{Characteristics of the the adversary}
\label{adv:characteristics}
  If $\mathcal{A}$ corrupts a party {\it actively}, it gains complete control over the party its input value, its random tape, its program and is free to send arbitrary messages on the behalf of the party, while also receiving all the messages sent to the party by other parties. The party is said to be {\it passively} corrupted when the adversary just gains the privilege to receive all the inputs, outputs and messages exchanged by it with other parties. We set up {\it adversary structures} to handle corruptions of parties as well as channels, passively as well as actively.

\subsection{Some notations relevant to secure multiparty computation}
  Let $\Pi$ be a multiparty protocol executed by a set of players $\mathcal{P}$. We define the {\it View} of a player as the set of inputs, random bits used by the player and all the messages received by the player during the execution of the protocol. Likewise, the {\it View} of adversary is the vector of views of the players corrupted by it. Further, the distribution of the views of the players/adversary is defined as the distribution of these views, generated from executing the multiparty protocol, taken over different random choices made by the players and the adversary. This distribution is defined for a vector of inputs given to the parties. Formally,

  Let multiparty computation protocol $\Pi$ be executed by a set of players $\mathcal{P}$. Variable $View_{p_j}^{\Pi,P,\mathcal{A}}(\overrightarrow{C},\overrightarrow{I})$ refers to the random variable denoting the view of $p_j$, when multiparty protocol $\Pi$ is executed by the set of players $P$ with input vector $\overrightarrow{I}$, when adversary $\mathcal{A}$ corrupts quadruplet $\overrightarrow{C}$. Correspondingly, the random variable $\overrightarrow{View}_{X}^{\Pi,P,\mathcal{A}}(\overrightarrow{C},\overrightarrow{I})$ denotes the vector of views of subset of players $X$, constituted from executing protocol $\Pi$ amongst set of parties $P$ with input vector $\overrightarrow{I}$. Distributions over these random variables are defined along the same lines and refereed to as ${\bf View}_{p_j}^{\Pi,P,\mathcal{A}}(\overrightarrow{C},\overrightarrow{I})$ and ${\overrightarrow{\bf View}}_{\overrightarrow{C}}^{\Pi,P,\mathcal{A},\overrightarrow{C}}(\overrightarrow{C},\overrightarrow{I})$, respectively.

\section{Formal definition of (unconditional) secure multiparty computation for the stand alone setting}
\label{sec:secmpc}
  Let $\Pi=(\Pi^1,\Pi^2)$ refer to a two phase multiparty protocol. Let $\mathcal{P}=\{P_1,P_2,\dots,P_n\}$ denote a set of parties. Let $\overrightarrow{y}=(y_1,y_2,\dots,y_n)$ denote the vector of input values, where $y_i = \overrightarrow{y}[i] \in \{0,1\}^{*}$ refers to the input value of the $i^{th}$ party $P_i$. We ascertain the correctness of commitment and computation phase of the protocol separately, while privacy property is ascertained for the entire protocol.

\subsection{Correctness of the input commitment phase}
\label{subsec:secmpc}
  Input commitment phase should possess the following properties:
\begin{enumerate}
\item{Honest parties commit to their intended input values:} All the honest parties are able to successfully commit to their initial input values, as long $\mathcal{A}$ corrupts less then or equal to $\lfloor \frac{n-1}{3} \rfloor$ parties.
\item{Commitment is binding:} Postfacto the termination of the Input commitment protocol, none of the parties are able to modify the committed values, irrespective of how the corrupted parties behave from here on.
\item{Non-malleability of commitment:} The adversary should not be able to commit to input values that are dependent/correlated with the input values committed by the honest parties, except for negligible probability. For information theoretic setting, this requirement can be seen as a finer aspect of privacy. If the adversary is successful in violating the non-malleability requirements, then it can distinguish input vectors which are more likely to have been committed by the honest parties, from those which are less likely, based on its view. Thus, this requirement is taken care of as a finer aspect of the privacy property.
\end{enumerate}

  Formally, input commitment protocol $\Pi^1$ is correct iff we can associate an $n$-variate function $reveal_{\Pi^1}(.,.,\dots,.)$ with it, which when applied to the transcripts of the parties generated by the execution of protocol $\Pi^1$, extracts the vector of input values committed to by the parties. Furthermore, for the honest parties the values reported by $reveal()$ are always same as the initial input values; and committed values of all of the parties are same even if arbitrary transcripts are substituted for the corrupted parties instead of the actual transcripts. If one can associate such a function $reveal()$ with $\Pi^1$, then protocol $\Pi^1$ is called correct. Lastly, the identity of the locations of transcripts of honest parties should not be relevant for this function, as long as $\lfloor \frac{2*n}{3} \rfloor + 1$ inputs of $reveal_{\Pi^1}$ are the true transcript values that were generated in the execution of the protocol.

\begin{remark}
  The domain of function $reveal()$ is the set of vectors of valid transcripts generated from the Input commitment phase and range is the set of vectors of values committed by the parties. Typically, the correctness property of Input commitment phase, is defined by requiring that there exists a corresponding algorithm REVEAL() using which the shared secret can be revealed/extracted by the parties, from the vector of their transcripts. Above $reveal()$ is a functional characterization of the VSS protocol, as opposed to an algorithmic one. Indeed, a constructive way of proving such a characterization i.e., proving that $reveal()$ exists is to demonstrate that there exists an algorithm by execution of which one may extract the secrets shared (i.e., input values committed) by the parties, from the transcripts.
\end{remark}

  Let $\overrightarrow{y},\Pi$ be as above. $\Pi^1(\mathcal{P},\overrightarrow{y},\overrightarrow{r^1},\mathcal{C},\mathcal{A})$ is used to refer to the vector of input values committed to by $\mathcal{P}$ on the execution of $\Pi^1$, starting with some vector of input values $\overrightarrow{y}$, randomness $\overrightarrow{r}$, when $\mathcal{A}$ corrupts subset $\mathcal{C} \subset \mathcal{P}$.

\begin{definition}
\label{twophaseprotocol}
  The input commitment phase $\Pi^1$ of protocol $\Pi$ is correct iff there exists an $n$-variate function $reveal_{\Pi^1}: \{\{0,1,\bot\}^*\}^n \rightarrow \{\{0,1\}^*\}^n$, which can be associated with $\Pi^1$ such that it satisfies the following properties.
  Let $\overrightarrow{Trans} = \Pi^1_{Trans}(\mathcal{P},\overrightarrow{y},\overrightarrow{r^1},\mathcal{C},\mathcal{A})$, denote the vector of transcripts of $\mathcal{P}$, generated by the execution of $\Pi^1$. Then,

\begin{enumerate}
\item{} Let $\overrightarrow{x} = reveal_{\Pi^1}(\overrightarrow{Trans})$. $\forall P_i \notin \mathcal{C}: \overrightarrow{x}[i] = \overrightarrow{y}[i]$ i.e., honest parties are able to corrupt their initial input values.
\item{} $\forall \overrightarrow{Trans'}: [(\forall P_i \notin \mathcal{C}: \overrightarrow{Trans'}[i] = \overrightarrow{Trans}[i]) \rightarrow (reveal_{\Pi^1}(\overrightarrow{Trans'}) = reveal_{\Pi^1}(\overrightarrow{Trans}))]$ i.e., the transcripts of the honest parties are sufficient to extract the values committed to by all the parties irrespective of the transcripts of the corrupted parties.
\end{enumerate}
  Further, $\Pi^1(\mathcal{P},\overrightarrow{y},\overrightarrow{r^1},\mathcal{C},\mathcal{A}) = reveal_{\Pi^1}(\overrightarrow{Trans})$, is used to refer to the vector of input values committed by $\Pi^1$.
\end{definition}

\subsection{Full definition of multiparty computation for stand alone setting, for vanilla model}
\label{subsec:full}
  A MPC protocol should possess two properties: Correctness and Privacy. Correctness of computation is defined in a straightforward manner. Our definition of privacy is based on the following understanding: Suppose a multiparty protocol is executed with some subset of corrupted parties. A view of the honest parties and corrupted parties is generated in the process. Based on this view, the adversary may try to make inference about the inputs of other honest parties. We require that an indistinguishable distribution of views of the adversary be computable just from the initial input values, committed input values and the output value of the corrupted parties. This amounts to arguing that the adversary can infer nothing more about the input values of honest parties, then what can be inferred from these values alone. More elaborate exposition is presented later in this section.

  Let $f:(\{0,1\}^{*})^n \rightarrow \{0,1\}^{*}$ be an $n$-ary functionality. Let $\overrightarrow{I} = (i_1,i_2,\dots,i_n)$ denote the vector of input values of the parties.

\begin{definition}
\label{App:traditional-MPC}
  Let $f$ be an $n$-ary function as defined above. Let $\Pi=(\Pi^1,\Pi^2)$ be a two phase multiparty protocol, as according to Definition \ref{twophaseprotocol}. Then, $\Pi$ {\bf securely evaluates} $f$ if the following conditions hold true $\forall \mathcal{C} \subset P$ of parties corrupted by $\mathcal{A}$, for which $|P - \mathcal{C}| \geq \lfloor \frac{2*n}{3} \rfloor + 1$:
\begin{enumerate}
\item{Correctness:}
  Let $\overrightarrow{x}$ refer to the vector of input values committed by the parties on execution of $\Pi^1$.\footnotetext{As defined above, the vector of input values committed by the parties is specified by vector $\overrightarrow{x} = \overrightarrow{\Pi^1(\mathcal{P},\overrightarrow{y},\overrightarrow{r^1},\mathcal{C},\mathcal{A})} = reveal_{\Pi^1}(\overrightarrow{Trans})$}. Then, for all honest parties $p_i \in P - \mathcal{C}$ the following holds true: $\Pi^2(\overrightarrow{x})_{p_i} = f(\overrightarrow{x})$.

\item{Privacy:} There exists simulators $(Sim_1,Sim)$, which take inputs the subset $\mathcal{C}$, $\overrightarrow{y}_{\mathcal{C}}$, $\overrightarrow{x}_{\mathcal{C}}$, $f(\overrightarrow{x})$, adversary program $\mathcal{A}$, and generates the distribution of views of the adversary $\mathcal{A}$, such that
\begin{enumerate}
\item{} \noindent $Sim^{\mathcal{A}}(\mathcal{C},\overrightarrow{y}_{\mathcal{C}},\overrightarrow{x}_{\mathcal{C}},f(\overrightarrow{x})) \approx \overrightarrow{View}_{\mathcal{C}}^{\Pi,\mathcal{A}}(\mathcal{C},\overrightarrow{y},\overrightarrow{x},f(\overrightarrow{x}))$
\item{} \noindent $Sim_1^{\mathcal{A}}(\mathcal{C},\overrightarrow{y}_{\mathcal{C}},\overrightarrow{x}_{\mathcal{C}}) \approx \overrightarrow{View}_{\mathcal{C}}^{\Pi^1,\mathcal{A}}(\mathcal{C},\overrightarrow{y},\overrightarrow{x})$
\end{enumerate}
\end{enumerate}
  for all feasible adversaries $\mathcal{A}$, for all functions $f$\footnotetext{The variable $\overrightarrow{View}_{\mathcal{C}}^{\Pi,\mathcal{A}}(\mathcal{C},\overrightarrow{x}_{\mathcal{C}},\overrightarrow{y}_{\mathcal{C}})$ is as defined at the beginning of the section, and used rather canonically in cryptographic literature}.
\end{definition}

\begin{remark}
\begin{enumerate}
\item{}
  Note that the simulator $Sim$ is given both $\overrightarrow{y}_{C}$ and $\overrightarrow{x}_{C}$ i.e., the initial input values and the input values committed by the corrupted parties and the output value $f(\overrightarrow{x})$. The simulator aborts and ignores those sessions when the corrupted parties commit to input values different then $\overrightarrow{x}_{C}$. For information theoretic regime there is no constraint on the running time of the simulator. The distribution is compared with the distribution of views of the adversary that are generated from the real execution of the protocol when the parties start with initial input values $\overrightarrow{y}$ and committed to $\overrightarrow{x}$. If the distributions generated from the two cases are proved indistinguishable, it amounts to saying that the adversary gains no more knowledge about the input values of other honest parties then what could be computationally derived just from the initial input values and the committed input values.

\item{} It may be possible that the corrupted parties are able to commit to values which are some how correlated to the input values of the honest parties. This should not be allowed. The requirement of simulator $Sim_1$ helps to achieve this {\bf non-malleability} requirement (which can be seen as a breach of privacy of honest parties) for the input commitment phase $\Pi^1$ in a clean manner. This is seen as follows. The definition of privacy implies that $Sim_1^{\mathcal{A}}(\mathcal{C},\overrightarrow{y}_{\mathcal{C}},\overrightarrow{x}_{\mathcal{C}}) \approx \overrightarrow{View}_{\mathcal{C}}^{\Pi^1,\mathcal{A}}(\mathcal{C},\overrightarrow{y},\overrightarrow{x})$. This in turn implies that for any two sets of vectors $\overrightarrow{x_1},\overrightarrow{y_1}$ and $\overrightarrow{x_2},\overrightarrow{y_2}$ for which $(\overrightarrow{x_1}_{\mathcal{C}},\overrightarrow{y_1}_{\mathcal{C}}) = (\overrightarrow{x_2}_{\mathcal{C}},\overrightarrow{y_2}_{\mathcal{C}})$ it is true that: $\overrightarrow{View}_{\mathcal{C}}^{\Pi^1,\mathcal{A}}(\mathcal{C},\overrightarrow{y_1},\overrightarrow{x_1}) \approx \overrightarrow{View}_{\mathcal{C}}^{\Pi^1,\mathcal{A}}(\mathcal{C},\overrightarrow{y_2},\overrightarrow{x_2})$. This is saying that, irrespective of what input vectors the honest parties start with, the views of the corrupted parties generated from the commitment phase, depend only on their own initial input values and the values committed by them. Thus, for every vector of input values committed by the adversary, each possible vector of input values of honest parties is equally likely.
\end{enumerate}
\end{remark}

\subsection{Exposition of the definition of privacy}
\label{subsubsec:privacy}
  Violating privacy property in the information theoretic setting amounts to an adversary inferring information about the input values of honest parties that it is not supposed to. Adversary can make any inference about the input values of other parties on the basis of its own view only. If it is shown that an indistinguishable distribution of views of the adversary can (always) be generated from a certain set of values, then it implies that the adversary can infer nothing more about the input values of honest parties than what can be information theoretically inferred on the basis of this set values, and the structure of the function $f$ only. Obviously, then the adversary cannot distinguish between the different vectors of input values of honest parties, which are (maximally-yet-equally) consistent with its own view, with any useful advantage. This is the understanding on which the definition of privacy is founded.

  Let us review the formal statement of the definition of privacy: There exists a simulator $Sim$ such that for all corruptions $\overrightarrow{C} \in \mathcal{T}$, vectors of initial input values $\overrightarrow{y}$ and committed input values $\overrightarrow{x}$ the following condition holds true:\\
 $Sim^{\mathcal{A}}(\mathcal{C},\overrightarrow{y}_{\mathcal{C}},\overrightarrow{x}_{\mathcal{C}},f(\overrightarrow{x})) \approx \overrightarrow{View}_{\mathcal{C}}^{\Pi,\mathcal{A}}(\mathcal{C},\overrightarrow{y},\overrightarrow{x},f(\overrightarrow{x}))$

  Now imagine a table consisting of the following three columns: Vector of committed input values of corrupted parties, vector of input values of honest parties (for honest parties the committed input values = initial input values, for the rest they may differ) and the output $f(\overrightarrow{x})$ that parties set out to compute. Fill up the rows of the table for all plausible choices of input values etc. that may be used in the multiparty computation. Different rows of the table must differ in at least one column entry. The goal of the adversary is to zoom on to the smallest subset of table entries which are (maximally-yet-equally) consistent with its view generated in real execution of the multi-party protocol. This set of compatible table entries correspond to the plausible vectors of input values that honest parties could have used in the computation. Table entries which are incompatible (or correspond to negligible probability of occurrence) with adversaries view are rejected. The guarantee of the definition of privacy is, that the set of non-rejected table entries must certainly include the following subset of entries: {\it The subset of rows which are compatible with the actual input values committed by the corrupted parties and the output value generated in real execution of the protocol.}
  Thus, fix the distribution of views of the adversary as $\approx Sim(\overrightarrow{I_i}, \overrightarrow{I_c}, Out)$ (where $\overrightarrow{I_i}$ is the vector of initial input values and $\overrightarrow{I_c}$ is the vector of committed input values of the corrupted and sacrificed parties) for some real execution of the multi-party protocol. Then the adversary cannot distinguish between any two distinct vectors $\overrightarrow{x}_1 \not \eq \overrightarrow{x}_2$ of input values, which the honest parties could have started with, for which $f(\overrightarrow{x}_1,\overrightarrow{I_c}) = f(\overrightarrow{x}_2,\overrightarrow{I_c}) = Out$, with any significant advantage.

\section{A new model for unconditional multiparty computation}
\label{interesting}
  Using the above approach for defining privacy we can give results for realizing almost everywhere secure computation (aka a.e.s.c.) on special families of incomplete networks for the case of general Byzantine corruptions, in a follow up work. Now, we consider a new intermediate model for multiparty computation when communication can face limited disruption and present results for it.

  In \cite{BCLPR}, authors consider a very weak computational model for MPC, when the adversary can fully control the communication between the parties. For such a feeble model, not much can be guaranteed because the adversary can complete disrupt any computation process. Yet theoretically this is an interesting model to consider and the results given in \cite{BCLPR} have an interesting flavor. It is shown that the adversaries strategy can be restricted to either message relaying or conducting independent executions with disjoint subsets of honest parties, and the only dependence that the adversary can achieve, between these executions, is by running these executions sequentially and choosing its input in second execution after receiving its input from the first.

  In \cite{V07}, \cite{GO08} consider a different model of MPC when only a partially connected network of secure channels is available to the parties, called a.e.s.c.. In contrast to \cite{BCLPR}, this model is theoretically and practically interesting for the case of information theoretic adversary only. On the other hand, although the adversary considered in \cite{BCLPR} is given too much power to assure any meaningful security guarantees for practical purposes, a.e.s.c. actually assures meaningful security guarantees to a large number of honest parties, if not all the honest parties (communistic settings are one example where they become relevant). Recall that the canonical guarantees of correctness and privacy has to be sometimes "sacrificed" for some honest parties that are surrounded by bad neighborhood of maliciously corrupted parties. Thus, depending on the statistical guarantees about the subset of maliciously corrupted parties, a.e.s.c. makes only statistical guarantees about the subsets of honest parties for which correctness and/or privacy properties are assured, for special families of incomplete networks. The reader may refer to \cite{GO08} for a thorough discussion on theoretical and practical reasons of studying a.e.s.c. on incomplete networks.

  Consider the following {\it intermediate model} (with respect to \cite{GO08} and \cite{BCLPR}), when an "information theoretic" adversary (which renders PKI based solutions irrelevant) can partially control a subset of communication channels in various ways. Depending on the actual subset of communication channels corrupted by the adversary and the type of corruptions of channels, not all honest parties may be able to achieve the guarantees of correctness and privacy just as in \cite{V07}, \cite{GO08}. For example, if a fairly large set of communication channels connecting an honest party are totally corrupted, then this party may not be able to commit to its intended input value or be guaranteed to receive its correct output value. On the other hand, if a large number of communication channels connecting to this honest party are authentic but eavesdroppable, then the privacy of its input value or that of its output cannot be guaranteed. Thus the canonical guarantees of correctness or privacy may have to be dropped for some of the honest parties. We give results for this intermediate model, in a.e.s.c. framework.
 
  From a practical point of view, also, many different types of man-in-middle attacks may be considered in realistic networks. Computer networks often face the threat of the following type of security attack. An intruder inserts itself between two communicating parties, both of which believe that they are talking to each other, while the attacker deletes, modifies or simply eavesdrops the messages exchanged between the parties. More generally, such an attacker can carry out the attack in a coordinated fashion and sabotage a large number of communication channels of the network. In particular, the attacker may use the messages received on one communication channel to modify or inject new messages on a different channel. At the same time interesting variations exist for the lower level protocols, for protecting integrity and privacy of the messages communicated over the channels. They give rise to interesting combinations of channels and man-in-middle attacks.

  We give results for this new model, called (Unconditional) secure multi-party computation with man-in-the-middle attacks, in a.e.s.c. framework. Realizing a.e.s.c. on incomplete networks, while handling the case of Byzantine corruptions, requires a more complex, cumbersome argument and has been presented in a follow up work.

\section{Modelling communication channels with a man-in-the-middle attacker}
\label{sec:modellingchannels}
  In this section we present an ideal functionality description for communication channels that captures different types of channel corruptions.

  A directed channel $e_{u,v}$ from party $p_u$ to party $p_v$ behaves as a {\it secure channel} if it is uncorrupted. The message received by the entity modelling the channel $e_{u,v}$ from party $p_u$ is received by party $p_v$ after a few rounds $r$ (a number which is physically hardwired in the channel). We consider a few different types of man-in-the-middle attacks by associating appropriate behavioral description with the communication channels as follows:
\begin{enumerate}
\item{Authenticated but eavesdroppable channel} models a commonly considered man-in-the-middle attack in which the adversary has the power to wiretap a communication line between two parties (say for example a telephone or a fiber optic cable) but cannot influence the integrity of the data communicated over it.
\item{Partially corrupted channel} Consider the following type of tampering of messages sent on the communication channel. The adversary gets a function $\alpha = f(x,r)$ of message $x$ sent over the communication channel. It corrupts $\alpha$ arbitrarily and forwards $\beta=func(\alpha)$ on the channel. The receiver receives the message $f^{-1}(\beta,r)$. This type of channel behavior models man-in-the-middle attack on a communication channel which may have some underlying mechanism to protect the confidentiality of the data sent on the channel. For example, the communication channel may hide the message by a one time pad before sending the message over the channel. When the message is received at the other end, the original one time pad is removed by XORRing the received message with sequence of random bits $r$, before it is passed to the party. In this sense, this type of corruption of the communication channel models the compliment of the authenticated but eavesdroppable channel. This is one variation which results in the case when the correctness property of some honest parties may be compromised, however privacy may or may not be preserved in different executions (for example, when the messages communicated over several such channels are corrupted so that the default value "d" is committed to by the relevant party which is known to all the rest of the parties).
\item{Fully tamperable channel} models the type of man-in-the-middle attack in which the adversary receives the message being sent on the channel as it is and gets enough number of rounds to corrupt the message arbitrarily before forwarding the message to the receiving party. This case considers the more traditionally studied flavor of man-in-the-middle attacks.
\end{enumerate}

  We assume a setting in which all the parties are synchronized with respect to a common clock. In every round, parties may engage in some computation after which they send some messages to other parties. The messages are exchanged by honest parties over the communication channel, which can be potentially corrupted in various ways. We assume that a message sent on the communication channel takes $r$ rounds for the receiver to receive and $r$ is publicly known constant. Furthermore, the adversary corrupting the message sent on the communication channel has limited flexibility on how it is allowed to interact with the communicated channel. For example, the adversary may be allowed only a limited number of rounds to corrupt the message sent on the channel and protocol may be constructed so a receiving party may discard untimely received messages (detecting that they were corrupted).

  Let $r>6$ be a publicly known constant. Let $\overrightarrow{C}$ denote a vector whose elements denote different types of channel and node corruptions as follows. The directed channel from $S$ to $R$ is referred to by $F_d^r(S,R,edge_{id})$ and the Ideal functionality for this directed channel from $S$ to $R$ is defined as follows:
\begin{definition}
\label{undirected}
  $F_d^r(S,R,edge_{id})$, denotes a directed communication channel from $S$ to $R$, with unique identity $edge_{id}$, synchronized with respect to a global clock, and executes as follows:
\begin{enumerate}
\item{$edge_{id} \in \overrightarrow{C}[2]$:} (Passively corrupted channel) If message $(S,R,mesg-id,m)$ is received from party $S$ in round $i$, then $F_d^r(S,R,edge_{id})$ records it and forwards the message $(S,R,mesg-id,m)$ to $\mathcal{A}$ in round $i+1$, and to party $R$ in round $i+r-1$.
\item{$edge_{id} \in \overrightarrow{C}[3]$:} (Partially corrupted channel)
\begin{enumerate}
\item{} If $(S,R,mesg-id,m)$ is received from $S$ in round $i$, $F_d^r(S,R,edge_{id})$ records the message, round number etc., chooses a sequence of random bits of appropriate length $r$, records it and forwards the message $(S,R,mesg-id,m \oplus r)$ to $\mathcal{A}$ in round $i+1$.
\item{} If $F_d^r(S,R,edge_{id})$ receives message $(S,R,mesg-id,m')$ to be sent to party $R$ from $\mathcal{A}$ in round $k$ for $k \leq i+r-2$ then continue, else drop the message. $F_d^r(S,R,edge_{id})$ checks the validity, time stamp etc of the message from previous records. $F_d^r(S,R,edge_{id})$ retrieves the sequence of random bits $r$ and forwards the message $(S,R,mesg-id, r \oplus m')$ to $R$ in round $i+r-1$.
\end{enumerate}
\item{$edge_{id} \in \overrightarrow{C}[4]$:} (Fully corrupted channel)
\begin{enumerate}
\item{} If $(S,R,mesg-id,m)$ is received from $S$ in round $i$, $F_d^r(S,R,edge_{id})$ records the message, round number etc. and forwards the message $(S,R,mesg-id,m)$ to $\mathcal{A}$ in round $i+1$.
\item{} In round $k$ for $k \leq i+r-2$, $F_d^r(S,R,edge_{id})$ receives the message $(S,R,mesg-id,m')$ to be sent to party $R$ from $\mathcal{A}$. $F_d^r(S,R,edge_{id})$ checks the validity, time stamp's etc of the message from previous records and forwards the message to $(S,R,mesg-id,m')$ to $R$ in round $i+r-1$.
\end{enumerate}
\item{Otherwise:} (Fully secure channel) If message $(S,R,mesg-id,m)$ is received from party $S$ in round $i$, $F_d^r(S,R,edge_{id})$ records it and forwards the message $(S,R,mesg-id,\bot,|m|)$ to $\mathcal{A}$ in round $i+1$, and message $(S,R,mesg-id,m)$ to party $R$ in round $i+r-1$.
\end{enumerate}
\end{definition}

\subsection{Adversary structure for the model and its characteristics}
\label{adv-characteristics}
  We set up adversary structures to handle the various combinations of corruptions of parties and channels.
\begin{definition}
\label{adversarystructure}
  Let $\mathcal{T} \subset \{(\mathcal{X}_p,\mathcal{X}_a,\mathcal{Y}_p,\mathcal{Y}_a,\mathcal{Y}_b,\mathcal{Y}_t) | \mathcal{X}_p,\mathcal{X}_a \subset V, \mathcal{Y}_p,\mathcal{Y}_a,\mathcal{Y}_b, \mathcal{Y}_t \subset V*V$, and $\mathcal{X}_p \bigcap \mathcal{X}_a = \phi, \mathcal{Y}_p \bigcap \mathcal{Y}_a \bigcap \mathcal{Y}_b \bigcap \mathcal{Y}_t = \phi$\}, where $\mathcal{X}_p$ denotes the subset of parties corrupted passively, $\mathcal{X}_p$ denotes the subset of parties corrupted actively, $\mathcal{Y}_p$ denotes the subset of channels corrupted passively, $\mathcal{Y}_a$, $\mathcal{Y}_b$ denote the subset of channels corrupted maliciously but partially and $\mathcal{Y}_t$ denotes the subset of channels corrupted maliciously and totally.
  An adversary $\mathcal{A}$ is called $\mathcal{T}$-restricted if all possible sextuplet of subset of parties corrupted passively and actively, subset of channels corrupted passively, and subset of channels corrupted partially and totally, by $\mathcal{A}$ belong to the set $\mathcal{T}$.
\end{definition}
  $\overrightarrow{C}$ is used to refer to a sextuplet of corruptions that belong to adversary structure $\mathcal{T}$. $\overrightarrow{C}$ is called {\it feasible corruption}, if there exists a subset of honest parties $H: |H| \geq \lfloor \frac{2n}{3} \rfloor + 1$, such that every two parties that belong to $H$ are connected via a secure channel. A {\it feasible adversary structure} $\mathcal{T}$ is defined along the same lines.

\subsection{Security implications for a given adversary structure}
\label{implications}
  In secure multi-party computation, the correctness property of an honest party is considered compromised if the honest party is not able to commit to its intended input value or it does not receive the correct output value. The privacy property is considered compromised if the adversary gets to learn even partial information about the input value of the honest party, that it is otherwise not supposed to. Depending on the type of channels by which an honest party may be connected to other participating parties, correctness or privacy may not be guaranteed to it. Depending on the specific element of adversary structure, we specify which participating honest parties are guaranteed the Correctness (and Privacy). Honest parties may have to sacrifice Correctness or Privacy depending on the type of communication channels they are connected to.
\begin{enumerate}
\item{} Every honest party that is part of a sub-clique of size at least $\lfloor \frac{2*n}{3} \rfloor + 1$ of passively corrupted or uncorrupted honest parties, which are connected to each other via authenticated-but-eavesdroppable or secure channels will be guaranteed the Correctness property.
\item{} Every honest party that is part of a sub-clique of size at least $\lfloor \frac{2*n}{3} \rfloor + 1$ of passively corrupted or uncorrupted honest parties, which are connected to each other via only secure channels will be guaranteed the Privacy property.
  Note that the condition does not allow the honest party to be connected via partially tamperable channels for which the adversary can corrupt the value being sent on the channel, without getting to learn the actual value being sent. This is because by manipulating data being sent on such channels and control of other corrupted parties, the adversary may falsify the honest party, get the party declared as "corrupt" (as happens in the input commitment phase of \cite{BGW88},\cite{CCD88} protocols) and them to commit to the default value "d". In this case, the default value is learned by all the participating parties and hence also the adversary. Thus, the privacy property of the party is compromised. Also, note that depending on how the adversary controls the communication channels, this situation may not always arise and honest party may be able to sometimes commit to its input value, which the adversary does not get to learn. However, whether or not this happens, depends on the actual execution of the protocol and choices made by the adversary. Since we cannot always guarantee the privacy of this honest party, we say that the privacy of such a party is sacrificed.
\end{enumerate}

\subsection{Correspondence between the adversary structure and (un)sacrificed parties}
  Based on the above conditions we formally describe a function $Comp$, whose domain is defined as the set of all (feasible) adversary structures and whose range is set of sets of tuplets of the form $(S_c,S_p)$, where $S_c$ is the subset of honest parties which are guaranteed the Correctness property and $S_p$ is the subset of honest parties which are guaranteed the Privacy property. The function $Comp$ will be used in the main security definitions for the new model. It is defined in terms of a function $func$ whose domain is the set of corruptible vectors (as discussed in Definition \ref{adversarystructure}) and whose range is the set of tuplets of form $(S_c,S_p)$, where semantic of $S_c$ and $S_p$ is as defined previously in this paragraph. Thus, we have $Comp(\mathcal{T}) = \{func(\overrightarrow{C_1}), func(\overrightarrow{C_2}), \dots, func(\overrightarrow{C_m})\}$, where $C_i$ is the $i^{th}$ feasible vector of corruptions as defined above. Function $func(\overrightarrow{C_i}) = (S_c,S_p)$ defines the subsets of honest parties which are guaranteed the Correctness and Privacy properties corresponding to a given vector of corruption.
\begin{enumerate}
\item{} $x \in S_c$ iff party $x$ belongs to a sub-clique $Sub$ of parties which are connected to each other via eavesdroppable-but-authenticated channels or secure channels and the size of this sub-clique $|Sub|$ is greater than $\lfloor \frac{2*n}{3} \rfloor + 1$.
\item{} $x \in S_p$ iff party $x$ belongs to a sub-clique $Sub$ of parties which are connected to each other via (only) secure channels and the size of this sub-clique $|Sub|$ is greater than $\lfloor \frac{2*n}{3} \rfloor + 1$.
\end{enumerate}

\subsection{Some technical subtleties in defining security for the new model}
\label{sec:understanding}
  We briefly argue why there is no satisfactory way of formulating definitions of security for the new model using the "Trusted third party" paradigm.

  For this model of multi-party computation(and likewise for a.e.s.c. on incomplete networks), we encounter scenarios for which the input value committed by (and the output value received by) a sacrificed honest party may or may not be influenced by the adversary, while the adversary may not always get to learn the actual value committed by the sacrificed honest party. We give an example to illustrate the issue: An honest party may be connected via several (partially) corrupted communication channels to other parties. The adversary can corrupt the message sent over these channels but is not able to view the actual message sent on them. Thus, depending on the choices made by it, the adversary could get the honest party declared as a 'fraud' and get it to commit to the default value "d". Clearly, all parties including the adversary gets to learn the input value committed by the party in such executions. However, in other executions of the protocol, the adversary may just pass the (hidden) message over the communication channel without disturbing it at all or modifying it slightly so that the input committed is a function of the original value and is not able to learn the input value committed by the party. Thus, whether or not the privacy of the honest party is compromised varies from one execution to another. Since it is not always guaranteed for this vector of corruption, $\overrightarrow{C}$, we say that the privacy of this honest party is "sacrificed".

  Depending on the type of corruptions of the communication channels connected to a given party, similar scenarios may arise when the input value of the sacrificed honest party may not be extractable from the adversaries view. Thus depending on the choice made by the adversary, such sacrificed parties may act neither as an honest parties, nor as actively controlled by the adversary. However, standard composition theorems for multiparty computation protocols are statements about the joint distribution of the views of the honest and corrupted parties and more then independent claims about Correctness and Privacy property for the parties. Since the sacrifice of only one of these properties for an honest party places it neither in the category of fully corrupted nor in the category of uncorrupted, there is no straightforward way to employ standard composition theorems for this model of secure multiparty computation.

   More generally, we find that there is no clean and satisfactory way to incorporate the interaction of sacrificed parties with the 'Ideal functionality' depending on the dynamic choices made by the adversary. In fact, it is not clear what output some of the sacrificed parties should get from the Ideal functionality, because in real execution of the protocol the output may be dynamically influenced by the adversary controlling multiple communication channels connected to the party. These issues have also been discussed in \cite{GO08}. Due to such technical issues, we are required to explore a different approach for formulating the definitions of security for the new model. Since we cannot utilize standard composition theorems, we restrict ourselves to the stand alone setting.

\subsection{Definition of security for the new model}
\label{subsec:incomplete}
  We propose definitions for the new model in almost everywhere secure computation framework. For this setting not all the honest parties are guaranteed to receive the correct output values or be able to preserve the privacy of their input values etc.. Honest parties for which this cannot be ensured, are said to have {\it sacrificed} these properties respectively.

  Let $\mathcal{T}$ be a feasible adversary structure, Section \ref{adversarystructure}. Recall the security implications of feasible adversary structure, where we study what subsets of honest parties can be guaranteed the Correctness and Privacy properties depending on the particular element of the adversary structure, $\overrightarrow{C}$. In particular, when the vector of parties and channels corrupted by the adversary is $\overrightarrow{C}$, we are required to guarantee correctness for subset $H_c$ and privacy for subset $H_p$, where $(H_c,H_p) = func(\overrightarrow{C})$. Globally, $Comp(\mathcal{T})$ is used to refer to the set of tuplets of subsets, namely $(H_c,H_p)$, which are guaranteed the Correctness and Privacy properties for the adversary structure $\mathcal{T}$. We now adapt the formal definitions given in previous section to include the notion of sacrifice for the new model.

\subsubsection{\bf Correctness of the input commitment phase}
  Let $(H_c,H_p) = func(\overrightarrow{\vec{C}})$, where the functions $func, Comp$ have been discussed in Subsection \ref{implications}. The input commitment protocol should have the following properties to be correct (1) All honest parties in subset $H_c$ should be able to commit to their initial input values, irrespective of the behavior of the corrupted parties (2) Binding: All parties are committed and the committed input values of all the parties are determined by the transcripts of the honest parties $H_c$ alone. Formally,
  Let $\mathcal{P}$ refer to a set of participating parties. Let $\mathcal{A}$ be an adversary restricted to adversary structure $\mathcal{T}$, Definition \ref{adversarystructure}. Let $\overrightarrow{C}$ refer to the sextuplet of parties and channels corrupted by the adversary $\mathcal{A}$ in a given execution.
\begin{definition}
\label{twophasealmosteverywhereprotocol}
  Let $\Pi=(\Pi^1,\Pi^2)$ be any {\it two phase} multiparty protocol. The input commitment phase $\Pi^1$ is ($\mathcal{T},Comp(\mathcal{T})$)-correct if there exists an $n$-variate function $reveal_{\Pi^1}: \{\{0,1,\bot\}^*\}^n \rightarrow \{\{0,1\}^*\}^n$, such that the following conditions hold true for all feasible $\overrightarrow{C}$ corrupted by $\mathcal{A}$.
  Let $\overrightarrow{Trans} = \Pi^1_{Trans}(\mathcal{P},\overrightarrow{y},\overrightarrow{r^1},\overrightarrow{C},\mathcal{A})$, denote the vector of transcripts of parties $\mathcal{P}$, generated by the execution of $\Pi^1$. Let $\overrightarrow{x} = reveal_{\Pi^1}(\overrightarrow{Trans})$. Let $(H_c,H_p) = Comp(\overrightarrow{C})$. The following condition holds for subset of (unsacrificed) honest parties $H_c$:
\begin{enumerate}
\item{Unsacrificed honest parties $H_c$ commit to their intended input values:} $\forall P_i \in H_c: x_i = y_i$, with probability greater than $1 - \mu(n)$, for some negligible function $\mu(.)$.
\item{Binding for all the parties:} $\forall \overrightarrow{Trans'}: (\forall P_i \in H_c: \overrightarrow{Trans'}[i] = \overrightarrow{Trans}[i]) \rightarrow (reveal_{\Pi^1}(\overrightarrow{Trans'}) = reveal_{\Pi^1}(\overrightarrow{Trans}))$.
\end{enumerate}
\end{definition} 

\subsubsection{Full definition of security for the new model} Correctness of the multi-party protocol follows by ensuring the correctness of the commitment phase and correctness of computation phase of the protocol. The requirements for the former have been formalized in the previous subsection. The requirements for the latter are relatively straightforward to formalize. First recall the feasible adversary structure $\mathcal{T}$ and the correspondence of adversary structure $\mathcal{T}$ to the unsacrificed honest parties, expressed by functions $func, Comp$ as discussed in Section \ref{adversarystructure}. We require that all honest parties belonging to subset $H_c$, where $(H_c,H_p)=func(\overrightarrow{C})$, should receive the correct output value $f(\overrightarrow{y})$, when the vector of input values committed to by the parties is $\overrightarrow{y}$. Privacy property of the multi-party protocol is formalized along the same lines as was done for the vanilla model in previous Section. Multi-party protocol is said to satisfy the privacy property, if there exists a simulator which can generate the distribution of the views of the adversary generated during real executions of the multi-party protocol, from the initial input values, committed input values of the parties belonging to subset $P - H_p$ (i.e., sacrificed and corrupted) and the output value alone. Formally,

\begin{definition}
\label{almosteverywhere}
  Let $\mathcal{T}, Comp,func$ and $f,\overrightarrow{y},P,\mathcal{A},\overrightarrow{C}$ be defined as before.
  Let $\Pi=(\Pi^1,\Pi^2)$ be a {\it two-phase} multiparty computation protocol, satisfying Definition \ref{twophasealmosteverywhereprotocol}, formalizing correctness of Input commitment phase $\Pi^1$. $\Pi$, $(\mathcal{T},Comp(\mathcal{T}))$-securely evaluates function $f$ if there exists a simulator $Sim$, such that for all $\overrightarrow{C} \in \mathcal{T}$ corrupted by $\mathcal{A}$, the following condition holds true for the subset of honest parties $H_c,H_p$, where $(H_c,H_p)=func(\overrightarrow{C})$:
\begin{enumerate}
\item{Correctness:} Let $\overrightarrow{x}$ be the vector of input values committed to by the parties, after the execution of input commitment phase $\Pi^1$\footnote{$\overrightarrow{x} \leftarrow \Pi^{1}(\overrightarrow{y},\overrightarrow{C},\mathcal{A},\overrightarrow{r^1})$, as defined above}. Then, for all $p_i \in H_c:$
  ${\Pi^2(\overrightarrow{x},\overrightarrow{C},\mathcal{A},\overrightarrow{r^2})}_{p_i} = f(\overrightarrow{x})$\footnote{The equality condition can be relaxed for the case of probabilistic function to $\approx$.}

\item{Privacy:} Simulator $Sim$, takes as input $\overrightarrow{C}$, $\overrightarrow{y}_{P - H_p}$, $\overrightarrow{x}_{P - H_p}$, output $f(\overrightarrow{x})$, adversary program $\mathcal{A}$ and generates a distribution of views of $\mathcal{A}$, such that:
  $Sim^{\mathcal{A}}(\overrightarrow{C}, \mathcal{T}, {\overrightarrow{y}}_{P - H_p}, \overrightarrow{x}_{P - H_p}, f(\overrightarrow{x})) \approx \overrightarrow{View}_{\overrightarrow{C}}^{\Pi,\mathcal{A}}(\overrightarrow{C},\overrightarrow{x},\overrightarrow{y},f(\overrightarrow{x}),\overrightarrow{r})$.
\end{enumerate}
  for all feasible adversary structures $\mathcal{T}$.
\end{definition}

  Note that in the above definition, the simulator $Sim$ is given the input values of some sacrificed honest parties. These values may not always be extractable from the actual view of the adversary. This issue has been addressed in the explanation of the above definition in the footnote below\footnotemark.

\footnotetext{
  Imagine a table consisting of the following three columns: Vector of committed input values of corrupted/sacrificed parties, vector of input values of unsacrificed honest parties (for honest parties the committed input values = initial input values, for the rest they may differ) and the output $f(\overrightarrow{x})$ that parties set out to compute. Fill up the table rows for all plausible values of multiparty computation. Different rows of the table must differ in at least one column entry. The goal of the adversary is to zoom on to the smallest subset of table entries which are (most-yet-equally) consistent with its view generated from real execution of the multi-party protocol. This set of compatible table entries correspond to the plausible vectors of input values that unsacrificed honest parties started with. Table entries which are incompatible (or correspond to negligible probability of occurrence) with adversaries view are rejected. The guarantee of the definition of privacy is the following. The subset of non-rejected table entries must certainly include the following subset of entries (as long as adversary is $\mathcal{T}$-restricted): {\it The subset of rows of the table which are compatible with the actual input values committed by the sacrificed honest parties, the corrupted parties and the output value generated in the real execution of the protocol.}

  Thus, fix the distribution of views of the adversary as $\approx Sim(\overrightarrow{I_i}, \overrightarrow{I_c}, Out)$ (where $\overrightarrow{I_i}$ is the vector of initial input values and $\overrightarrow{I_c}$ is the vector of committed input values of the corrupted and sacrificed parties) for some real execution of the multi-party protocol. Then, the adversary cannot distinguish between any two vectors $\overrightarrow{x},\overrightarrow{y}$ of initial input values, with which unsacrificed honest parties could have started with, as long as it is true that $f(\overrightarrow{x},\overrightarrow{I_c}) = f(\overrightarrow{y},\overrightarrow{I_c}) = Out$.
}

  We have the following result for the new model.
\begin{theorem}
  There exists a multiparty computation protocol $\Pi'$ that $\mathcal{T}$-securely evaluates function $f$ according to Definition \ref{almosteverywhere}, for all feasible adversary structures $\mathcal{T}$.
\end{theorem}

  Multiparty computation protocols like \cite{BGW88}, \cite{CCD88} etc. are known to be resilient against a computationally unbounded adversary which can corrupt at most $\lfloor \frac{n-1}{3} \rfloor$ parties. Thus, we have the following well known claim for the stand alone setting.
 
\begin{theorem}
  There exists a multiparty computation protocol $\Pi$ that securely evaluates function $f$ as according to Definition \ref{App:traditional-MPC}.
\label{traditionalclaim}
\end{theorem}

  Using the above Theorem \ref{traditionalclaim}, we show how to realize almost everywhere secure multiparty computation when a subset of communication channels can also be corrupted.

\begin{theorem}
  There exists a multiparty computation protocol $\Pi'$ that $\mathcal{T}$-securely evaluates function $f$ according to Definition \ref{almosteverywhere}, for all feasible adversary structures $\mathcal{T}$.
\label{edgeCorruptMPC}
\end{theorem}

\begin{proof}
  Let $\Pi$ be an (unconditionally) secure multiparty protocol from Theorem \ref{traditionalclaim}, for the vanilla setting. The underlying assumption for Theorem \ref{traditionalclaim} is that it takes one round for a message sent on the secure channel to reach the party at the other end. Thus, a trivial variation is to consider an adaptation of the multiparty protocol $\Pi$ when the underlying communication channels take $r$ rounds to deliver the messages sent on it (Thus parties engage in computation only at the $0 mod r^{th}$ round). This is achieved by blowing each single round of multiparty protocol $\Pi$ to a slot of $r$ rounds, such that the parties are now activated only at the beginning of each slot (i.e., in corresponding rounds $0 mod r$). Let $\Pi'$ be this adapted multiparty protocol that satisfies the Theorem \ref{traditionalclaim} for this set up. Further modify $\Pi'$ so that each ordered pair of parties (of the $n*n-1$ ordered pairs in all) is provided a unique non-overlapping slot for transmission, arranged in an arbitrary but predetermined order. Thus, each slot of $\Pi'$ is now replaced with a block of $n*(n-1)$ non-overlapping super-rounds, each of which itself consists of $r$ rounds. Let $\Pi''$ be this resulting protocol. In protocol $\Pi''$ the first super-round is allotted for party $P_1$ to transmit any message to party $P_2$, the second super-round is allotted for party $P_1$ to transmit any message to party $P_3$ and so on and so forth. The entire sequence of $n*(n-1)$ such super-rounds constitutes a block of super-rounds of protocol $\Pi''$.

  We shall show that protocol $\Pi''$, $(\mathcal{T},Comp(\mathcal{T}))$-securely evaluates function $f$ as according to Definition \ref{almosteverywhere}, for all feasible adversary structures $\mathcal{T}$. The proof is a reduction argument which proceeds in two stages. In the first stage, we claim that Theorem \ref{edgeCorruptMPC} holds true for all feasible adversary structures, that have a special type of adversary structure, if Theorem \ref{traditionalclaim} is true. In the second stage, we shall show that if Theorem \ref{edgeCorruptMPC} holds true for all {\it special} feasible adversary structures, then it holds true for all feasible adversary structures.

  The special feasible adversary structures are feasible adversary structures which have no channel corruptions of type 3 and type 4 i.e., for these feasible adversary structures $\overrightarrow{C}[3]=\overrightarrow{C}[4]=\phi$. There is a straightforward reduction from Theorem \ref{edgeCorruptMPC}, for the case of adversaries restricted to such feasible adversary structures, to Theorem \ref{traditionalclaim}. Intuitively, this is straightforward because corruption of communication channels of type $3$ and $4$ do not need to be handled and passive corruption of channels is taken care of by passive corruption of corresponding parties whenever necessary (when an honest party is connected to at least $\frac{n-1}{3}$ other parties via passively corrupted channels or the parties are themselves passively/actively corrupted).

  We now focus on the second stage reduction, in which we will show that for every $\mathcal{T}$-restricted adversary $\mathcal{A}'$ that attacks the execution of protocol $\Pi''$, there exists a corresponding adversary $\mathcal{A}$, that attacks the execution of protocol $\Pi''$, but is restricted to a corresponding special feasible adversary structure described above, for which Theorem \ref{edgeCorruptMPC} holds true. We now describe the adversary structure for $\mathcal{A}$.
 
  Let the adversary $\mathcal{A}'$ corrupt a feasible sextuplet $\overrightarrow{C}' \in \mathcal{T}$. Since $\overrightarrow{C}'$ is feasible, so there exist subsets of honest parties $H_c,H_p \subset P$, $(H_c,H_p) = func(\overrightarrow{C}')$ such that $|H_c|,|H_p| \geq \lfloor \frac{2*n}{3} \rfloor + 1$ for which we are required to satisfy the Correctness and Privacy properties as according to Definition \ref{almosteverywhere}.

  The strategy of $\mathcal{A}$ is as follows. $\mathcal{A}$ corrupts $\overrightarrow{C}$, such that $\overrightarrow{C}[0] = H_c - H_p$ (passively corrupted) and $\overrightarrow{C}[1] = P - H_c$ (maliciously corrupted). Furthermore, $\forall (p_u,p_v): p_u \& p_v \notin H_p \longrightarrow (p_u,p_v) \in \overrightarrow{C}[2]$ i.e., all channels outside subset $H_p$ are passively corrupted. This is the only type of corruption we consider for $\mathcal{A}$ i.e., $\overrightarrow{C}[3]=\overrightarrow{C}[4]=\phi$ and the rest of the channels are inside $H_p$ and are secure. Thus, from our construction of $\overrightarrow{C}$, it is easy to verify that $func(\overrightarrow{C}) = (H_c,H_p)$ (from the definition of $func$ in Subsection \ref{implications}).

  $\mathcal{A}$ will attack the execution of protocol $\Pi''$ for this corruption, while internally simulating the view of $\mathcal{A}'$ perfectly. We shall show that the parties belonging to subset $H_c$ and $\mathcal{A}'$ cannot distinguish between these two scenarios, namely when they are participating in Case A) $\mathcal{A}'$ attacking execution of protocol $\Pi''$, while corrupting $\overrightarrow{C}$ or Case B) $\mathcal{A}$ attacking execution protocol $\Pi''$, while corrupting $\overrightarrow{C}$ and internally simulating $\overrightarrow{A}'$. Correctness and Privacy of protocol $\Pi''$ against $\mathcal{A}'$ would then follow. $\mathcal{A}$ acts as follows:
\begin{enumerate}
\item{For parties belonging to $\overrightarrow{C}[1]$:} $\mathcal{A}$ internally simulates $\mathcal{A}'$ for the maliciously corrupted parties belonging to the subset $\overrightarrow{C}[0]$. The messages to be delivered by parties belonging to $\overrightarrow{C}[0]$ to other parties and vice verse are first passed through the simulations of corresponding communication channels, which may or may not be (partially) controlled by $\mathcal{A}'$. Note that some of the parties belonging to $\overrightarrow{C}[1]$ are not maliciously corrupted by $\mathcal{A}'$ but behave corrupted because there are too many corrupted communication channels connected to them. For these case, also $\mathcal{A}$ carries out the appropriate simulation by first simulating the protocol for honest party, then simulating the communication channel on which this message is supposedly to compute the message which is finally sent by $\mathcal{A}$.
\item{For parties belonging to $\overrightarrow{C}[0]$:} $\mathcal{A}$ internally simulates the role for honest parties for parties belonging to $\overrightarrow{C}[0]$ while also appropriately simulating the view of $\mathcal{A}'$.
\item{For communication channels:} The messages sent by the parties to other parties and received by the parties from other parties, may be (partially/fully) corrupted when they are sent on the communication channels. These communication channels are internally simulated by $\mathcal{A}$ to correspondingly simulate the view of $\mathcal{A}'$, before the possibly modified message is transmitted to other parties Or are internally simulated to correspondingly constitute the view of $\mathcal{A}'$ after the message has been received by the party via the secure/passively corrupted channel.
\end{enumerate}

  We claim that the following two conditions are simultaneously true after every super-round of execution of protocol $\Pi''$ for the two settings: Case A) when $\mathcal{A}'$ attacks execution of protocol $\Pi''$, while corrupting $\vec{C}'$: Case B) when $\mathcal{A}$ attacks execution of protocol $\Pi''$, while internally simulating $\mathcal{A}'$ as described above and corrupting $\overrightarrow{C}$.
\begin{enumerate}
\item{} The distribution of the views of the honest/semi-honest parties in $H_c$ are indistinguishable after every super-round.
\item{} The distribution of the views of $\mathcal{A}'$ generated from Case (A) and Case (B) are indistinguishable after every super-round.
\end{enumerate}

  It is easy to verify from the construction of $\mathcal{A}'$, described above, that if these two conditions are simultaneously true after super-round $j$ for case (A) and case (B), then they are simultaneously true after super-round $j+1$ etc.
 
  Hence the distribution of transcripts of the parties at the end of the Input commitment phase are indistinguishable. In particular, the same function $reveal_{\Pi''}$, from case (A), also possesses the requisite characterization of the Input commitment phase of protocol $\Pi''$ being attacked by $\mathcal{A}'$. Now conditioned to the fact that the input values committed to by all the parties are the same for two executions corresponding to case (A) and case (B), the distribution of transcripts of the honest parties belonging to the subset $H_c$ are also indistinguishable after the last round of the protocols $\Pi'$ and $\Pi''$, which implies that the output values for deterministic functions (and the distribution of output values for the case of probabilistic functions) of the honest parties are also same. The {\bf Correctness} property of protocol $\Pi''$ when adversary $\mathcal{A}'$, corrupting $\overrightarrow{C}'$, attacks its execution follows from {\bf Correctness} property of protocol $\Pi''$ when adversary $\mathcal{A}$, corrupting $\overrightarrow{C}$, attacks its execution.

\noindent{Privacy:}
  The proof of {\bf Privacy} of protocol $\Pi''$, when $\mathcal{A}'$ attacks its execution, just involves skillfully transferring the above work.

  For the privacy condition it is enough to demonstrate an appropriate simulator $Sim'$ which satisfies the conditions as according to Definition \ref{almosteverywhere}. The simulator $Sim'$ is given the initial input values and the input values committed to by the parties belonging to subset $P-H_p$ besides the output value and is required to generate distribution of views of the adversary $\mathcal{A}'$ that is indistinguishable from the ones generated in real execution of the protocol $\Pi''$.

  Now recall the construction of subset of honest parties $H_p$, for which the Privacy property is guaranteed. Each honest party $P_i \in H_p$ is connected via secure channels to at least $\lfloor \frac{2*n}{3} \rfloor$ honest parties which are connected to each other via secure channels only. It is easy to verify that by our construction of $\mathcal{A}$ and the definition of functions $Comp$ and $func$ for feasible adversary structures: the subset of parties $H_p$ for Case (A) (when $\mathcal{A}'$ attacks the protocol execution), is the same as the subset of parties $H_p$ for Case (B) (when $\mathcal{A}$ attacks the protocol execution).

  There exists a simulator $Sim$ that generates the distribution of views of adversary $\mathcal{A}$ which is indistinguishable from the distribution of the views of these parties generated from real execution of the protocol, for Case (B) by definition of privacy property. Simulator $Sim$ is given initial input values, committed input values for parties belonging to subset $P-H_p$ besides the output value. The same simulator $Sim \Leftrightarrow \mathcal{A}$ can be invoked for Case (A) as well.

  For case (A), we know from above that $\mathcal{A}$ internally simulates $\mathcal{A}'$ i.e. $Sim \Leftrightarrow \mathcal{A} \Leftrightarrow \mathcal{A}'$. Now let $Sim' = Sim \Leftrightarrow \mathcal{A}$ who in turn interacts with $\mathcal{A}'$ to generate the distribution of the views of $\mathcal{A}'$ i.e.,  $Sim' \Leftrightarrow \mathcal{A}'$. If the distributions of views of $\mathcal{A}'$ generated by the above simulator $Sim'$, and the distribution of views of $\mathcal{A}'$ generated during the real execution of protocol $\Pi'$ are distinguishable, then it translates to distinguishability in the distribution of views of $\mathcal{A}$ generated from the real execution of the protocol, from the distribution of views of $\mathcal{A}$ generated by the simulator $Sim$, contradicting the privacy property for protocol $\Pi''$ for Case (B).
\end{proof}

\begin{remark}
  However, simulations when the committed input values for corrupted and sacrificed parties are different then the ones given to the simulator (What values have been committed to by the corrupted and sacrificed parties in the commit phase is verified at the end of the first phase of the protocol as the simulator can compute the committed input values for these parties from its own view) are discarded and are only required to produce the simulation for the case when the input values committed to by these (corrupted and sacrificed) parties is the one given as only that case is required to correspond to the output values - given to the simulator. Therefore, we only consider the simulated output distribution of the views of the adversary $\mathcal{A}'$ conditioned to the above fact.
\end{remark}

\hspace{0.5in}
{
\small
\bibliographystyle{alpha}
\newcommand{\etalchar}[1]{$^{#1}$}

}

\appendix

\section{An alternate exposition of the definition of privacy of new model}
\label{sec:expose}
  We adapt the underlying concept of simulation in Zero-Knowledge protocols to formulate the definition of the privacy property for the stand alone setting, for the vanilla model. Subsequently, the notion of "sacrifice" of honest parties is incorporated in the definition, for the new model. This conceptual exposition has been included here for the sake of completeness from \cite{V10b}.

  We shall first examine the role of simulator in several different definitions of security of two party or multiparty protocols and what demonstrating a simulator for these applications is supposedly understood. We then present a meta-definition which captures the bare bone "skeletal" of these definitions. We then explain how the understanding subsumed in the meta-definition is employed to define privacy property of these protocols and also in this work.

  The celebrated work on Zero Knowledge Proofs introduced the fundamental notion of knowledge complexity and defined $0$-Knowledge Proofs (ZKP)\footnote{One may want to keep in mind the following sentence from page 295 (second last paragraph) of the historic paper \cite{GMR85} while formulating the notion of knowledge, "With this in mind we would like to derive an upper bound (expressed in bits) for the amount of knowledge that a polynomially bounded machine can extract from a communication. Further review the definition for $L \in KC(f(n))$}. Loosely, a protocol is called a $0$-KP if the verifier does not gain anything from interacting with the prover which it could not have generated by itself, except for one bit i.e., validity of some statement. This is formalized as follows: A protocol is ZKP iff there exists a PPT simulator which given the inputs available to both the parties, the verifier's program and any auxiliary values with the verifier, can generate a distribution of transcripts of the verifier which is indistinguishable from the distribution of transcripts generated in real execution of the protocol. Let us examine this closely by considering an example.

  Consider a ZKP in which the verifier is given an auxiliary input $Aux$ but has limited space (i.e., space bounded). The (cheating) verifier tries to use the auxiliary input in order to extricate some extra knowledge while executing the protocol with the prover. As a result a transcript $T$ is generated with the verifier. Afterwards, due to space constraints the verifier deletes the auxiliary input $Aux$. Clearly, now the verifier does not seem to be able to generate an indistinguishable looking transcript of the verifier, because the verifier does not have the auxiliary input $Aux$ with it, which the simulator takes as input. Does the same transcript $T$ which was earlier considered to convey $0$-knowledge, now represent anything more to the verifier ? Why or why not? How much knowledge is contained in the transcript (of and) for the verifier? A statement that is consistently true both before and after the verifier deletes the auxiliary input $Aux$ from its tape is: Whatever knowledge the verifier could (PPT) computationally derive from the transcript $T$, can also be computationally derived just from the initial input value which is given to both parties, the verifier's program and the auxiliary input value $Aux$. When the verifier also possessed the auxiliary input then this could be interpreted to mean that $T$ contained $0$-knowledge for the verifier. Similar understanding is used to define the privacy property of multiparty protocols proved secure via simulation paradigm: An ideal process and real process are described. In the ideal process the power of the adversary looks "somewhat" curtailed and parties are given access to an ideal functionality, while the simulator simulates the protocol execution and generates views of all the parties. Distribution of these views of the adversary generated from the two processes are proved to be indistinguishable for the two cases to claim the privacy property of the protocol. But what is really implied in inferring the privacy property from such a proof, as in \cite{GMW87} ? It is that the view of the adversary (for the ideal process case) could be generated by a (PPTM or just TM) simulator which is given access only to some input values and output values of the corrupted parties. Nevertheless, aside from these input and output values there are many other messages that are part of the view of the adversary even for the case of ideal process. So the underlying understanding is: The adversary cannot derive any more information about the input values of the honest parties, then what can be (PPT) computationally (or information theoretically whatever the case may be) derived just from its own input and output values. This is interpreted as the Privacy property of the protocol.

  Towards this end, we intend to capture the essence of demonstrating a simulator in definitions of ZKP and some other definitions of multiparty protocols (like the original definition in \cite{GMW87}). The following meta-definition of simulator captures the essential skeletal of these definitions of security: {\it A simulator is hypothetical mental construct which is used to prove properties that should exist about the relations between input values, intermediate values and output values generated by the execution of a multiparty protocol with a given adversary.}

  Let us see how we arrive at this meta-definition. First see that the distribution of transcripts generated by simulator $Sim$ by interacting with verifier's program $V$ can be produced just by a single Turing Machine $U$ which is given the following auxiliary inputs: A string of bits encoding the program of the simulator $Sim$, a string of bits encoding the program of the verifier $Ver$. The Turing machine $U$ takes as input, the input value $I$ and the auxiliary value of the verifier $Aux$. $U$ is also given a random tape. $U$ simulates internally the interaction between the simulator and the verifier by interpreting the strings $Sim$ and $Ver$ as two procedures which have separate tapes/memories for performing read and write, but who are also given a common shared memory corresponding to the interactive tapes for communication. $U$ detects when the simulation has failed and verifier needs to be rewound to an earlier state and does this when necessary by maintaining a stack. Finally, $U$ outputs a distribution of transcripts. It is easily seen that $U$ runs in time polynomial in the running times of the original simulator $Sim$ and the verifier $Ver$ and produces the output in one shot without interactive computation of any kind.

  The output of Turing Machine $U$ on input $I, Aux$ is the requisite distribution $D^I_u$ i.e., $U(I, Aux)_r = D^I_u \approx P \leftrightarrow V(I, Aux)$. Looked another way $U$ is a Turing Machine which is just computing the value of a probabilistic function $func(.,.)$ whose domain is all possible tuplets that correspond to values that can be assigned to $I, Aux$ and range is a distribution of transcripts of the verifier. Furthermore, $func(I, Aux)_r \approx P \leftrightarrow V(I, Aux)$. The property of function $func(.,.)$ that it is PPT computable is interpreted to mean that the verifier does not derive anything more computationally from interacting with the prover, then what could be derived by the verifier itself. We emphasize the point here that the verifier may or may not have access to the simulator or the auxiliary values to participate in the simulation at the time of simulation. However, it is the property of the function func(.,.) i.e., PPT computability of $func(.,.)$ that matters here and which we care about. In other words, the simulator is merely an "abstract" construct which has nothing to do with what may or may not be achievable/available in reality and is used only to demonstrate how the initial input, auxiliary input, string encoding verifier's program relate to the intermediate and output values i.e., the transcripts generated in the execution of the protocol. What is achieved by demonstrating such a simulator is a proof of some properties of the "existing" relations between the different values possessed by possibly different parties. This methodology is "constructive" (we demonstrate an algorithm). The interesting property that the function $func(.,.)$ possesses from our perspective is that it is PPT computable.

  This understanding is extended to the case of multi-party protocols in a straightforward manner.

\noindent {\bf More specific comments about the definition of privacy in this work:}
  For the information theoretic case, violating the privacy property of the multi-party protocol can be reduced to the adversary distinguishing between plausible input vectors of unsacrificed honest parties on the basis of its own view. If it can be shown that the view of the adversary can be (indistinguishably) generated using only some vectors of input values of corrupted, sacrificed partie and output values, we are done. What we infer from this proof is that the adversary has gained nothing more about the input vectors of unsacrificed honest parties than what could be information theoretically inferred using these values alone\footnote{We give an example of a typical case that arises in our analysis, that more clearly justifies the choice of definition on our work. If an honest party is connected via tamperable-but-private channels to several ($> \lfloor \frac{n}{3}+1 \rfloor$) other parties, then the message communicated on these channels can be corrupted and in fact the input commitment phase for this party can completely fail. In this case, the default value "d" is committed on behalf of this party, which is known by all other parties. Depending on the choices made by the adversary this may not always happen and the honest party may sometimes be able to commit to its initial input value, without this value being divulged to the adversary. Thus, whether the privacy property of this otherwise honest party is compromised or not, depends on the dynamic choices made by the adversary. But, if the privacy property for a party is ensured sometimes and not ensured at other times, then it is "sacrificed" i.e., "not guaranteed". Such cases occur in our analysis in this work and justify why the simulator is given the input values of these "sacrificed" honest parties always - hence this definition}.

  Under the light of the above discussion, let us review the formal definition of privacy in this work: There exists a simulator $Sim$ such that for all feasible corruptions, initial input values $\overrightarrow{y}$ and committed input values $\overrightarrow{x}$ the following condition holds true:\\
  $Sim^{\mathcal{A}}(\overrightarrow{C}, \mathcal{T}, N, {\overrightarrow{y}}_{P-H}, \overrightarrow{x}_{P-H}, f(\overrightarrow{x})) \approx \overrightarrow{View}_{\overrightarrow{C}}^{\Pi,\mathcal{A}}(\overrightarrow{C},\overrightarrow{x},\overrightarrow{y},f(\overrightarrow{x}),\overrightarrow{r})$.
  Consider another set of vectors $\overrightarrow{x'}$ and $\overrightarrow{y'}$, such that $\overrightarrow{y'}_{P-H} = \overrightarrow{y}_{P-H}$, $\overrightarrow{x'}_{P-H} = \overrightarrow{x}_{P-H}$ (i.e., values corresponding to sacrificed and corrupted parties) and $f(\overrightarrow{x}) = f(\overrightarrow{x'})$. For these values privacy property implies that:\\
  $Sim^{\mathcal{A}}(\overrightarrow{C}, \mathcal{T}, N, {\overrightarrow{y'}}_{P-H}, \overrightarrow{x'}_{P-H}, f(\overrightarrow{x'})) \approx \overrightarrow{View}_{\overrightarrow{C}}^{\Pi,\mathcal{A}}(\overrightarrow{C},\overrightarrow{x'},\overrightarrow{y'}, f(\overrightarrow{x'}) = f(\overrightarrow{x}), \overrightarrow{r})$.

  Now observe that the inputs of the simulator $Sim$ are same for both the cases as $\overrightarrow{y}_{P-H} = \overrightarrow{y'}_{P-H}$, $\overrightarrow{x}_{P-H} = \overrightarrow{x'}_{P-H}$ and $f(\overrightarrow{x}) = f(\overrightarrow{x'})$, which implies that:\\
  $\overrightarrow{View}_{\overrightarrow{C}}^{\Pi,\mathcal{A}}(\overrightarrow{C},\overrightarrow{x},\overrightarrow{y},f(\overrightarrow{x}),\overrightarrow{r}) \approx \overrightarrow{View}_{\overrightarrow{C}}^{\Pi,\mathcal{A}}(\overrightarrow{C},\overrightarrow{x'},\overrightarrow{y'},f(\overrightarrow{x}),\overrightarrow{r})$ i.e., {\it As long as the initial input values, committed input values of the sacrificed honest parties and corrupted parties and the final output value are same, the distribution of views of the adversary are also same.} i.e., the adversary cannot distinguish between the two set of input values of the unsacrificed honest parties used in computation with any advantage.

\end{document}